\documentclass[letterpaper]{article}
\usepackage{aaai}
\usepackage{times}
\usepackage{helvet}
\usepackage{courier}
\usepackage{booktabs}
\usepackage{graphicx}
\usepackage{amssymb}
\usepackage{epstopdf}
\usepackage{amsmath}
\usepackage{url}

\newcommand{\pview}[1]{\ensuremath{{p_{#1}}}}

\usepackage{caption}
\usepackage{subcaption}

\usepackage[]{color-edits}
\addauthor{gs}{blue}

\title{Popularity and Quality in Social News Aggregators: A Study of Reddit and Hacker News}
\author{Greg Stoddard \\
Northwestern University
}

\begin{document}

\nocopyright

\maketitle

\begin{abstract}

In this paper we seek to understand the relationship between the online popularity of an article and its intrinsic quality. Prior experimental work suggests that the relationship between quality and popularity can be very distorted due to factors like social influence bias and inequality in visibility. We conduct a study of popularity on two different social news aggregators, Reddit and Hacker News. We define quality as the relative number of votes an article would have received if each article was shown, in a bias-free way, to an equal number of users. We propose a simple poisson regression method to estimate this quality metric from time-series voting data. We validate our methods on data from Reddit and Hacker News, as well the experimental data from prior work. This method works well even though the collected data is subject to common social media biases. Using these estimates, we find that popularity on Reddit and Hacker News is a stronger reflection of intrinsic quality than expected.  

\end{abstract}

\section{Introduction}

One of the many narratives surrounding the growth of social media is that our systems for liking, retweeting, voting, and sharing are giving rise to a digital democracy of content. As the narrative goes, virality enabled  ``Gangnam Style'' to dominate international audiences, helped the Ice Bucket challenge raise millions of dollars for ALS research, and we now interpret trending topics on Twitter as a signal of societal importance \cite{gillespie2014}.  There's a considerable amount of academic work that interrogates this narrative by delving deeply into understanding the properties of virality. For example, scholars have studied the propagation and correction of rumors\cite{friggeri2014rumor}, the role of influential users in spreading information \cite{bakshy2012role}, or whether information actually diffuses in a viral way at all \cite{goel2012structure}. 

Although many papers hint at it, there are few papers that directly address a basic question: do these systems promote the best content? Does this ``digital democracy'' actually work? As a thought experiment, imagine polling a large population of people and asking them to rate every music video uploaded to Youtube in 2012. Would ``Gangnam Style'', the most watched video on Youtube, still come out on top? 

Evidence from the MusicLab experiment of \citeauthor{salganik2006experimental} \shortcite{salganik2006experimental,salganik2008leading} suggests that it might not. In this experiment, the authors set up a website where users could listen to and download songs from unknown artists. When visiting the site, participants were randomly assigned to a ``world'' and presented a list of songs that were ranked by the number of downloads the song had in that world. This design let the authors observe the evolution of popularity of the same song across different worlds. They also included one world in which songs were ranked randomly. The number of song downloads in this control world served as a measure of intrinsic song quality. 

They found that the popularity of a song could vary wildly across worlds; songs with the largest share of downloads in one world went relatively ignored in another one. Higher quality songs were more popular on average, but there was a large variance in popularity for all but the best and worst songs. This variance was caused by a rich-get-richer effect. Songs with more downloads were ranked higher in the list and were more likely to be sampled by future listeners. Furthermore, participants were able to see the number of current downloads each song had, and were more likely to sample songs with a higher number of downloads. In the presence of such effects, the authors conclude, popularity is a noisy and distorted measure of quality.  

\textbf{Present Work}

What do these results imply about the relationship between intrinsic quality and popularity on real world socio-technical systems?  Facebook, Twitter, etc all have a rich-get-richer phenomenon because posts with more likes, retweets, and views are more visible, on average, than their less popular counterparts. Does this imply that there's a distorted relationship between quality and popularity on these platforms? Unfortunately we do not have the ability to run randomized experiments on these platforms, so the main challenge of answering this question is developing a metric of quality that can be estimated from observational popularity data. 

In this paper we argue that social news aggregators provide a good setting to study the relationship between popularity and article quality. We conduct our study on two aggregators, Reddit and Hacker News. Reddit is a popular site where users submit links to content from around the web, and other users vote and comment on those links. Hacker News is an aggregator dedicated to programming and technology-related issues but is otherwise similar in structure. Reddit received approximately 450 million page views in December 2014, while Hacker News received approximately 3.25 million. 

These aggregators have several properties that facilitate disentangling observed popularity from inherent quality. The first property is the rich-get-richer effect is easier to measure on Reddit and Hacker News because content visibility is easier to quantify. The interface of each site is a simple, non-personalized list of links\footnote{Reddit is actually lightly personalized, see section \ref{sec:data} for more details.}, so the observed article ranking is (approximately) the same for all users. Due to the similarities in UI, estimating visibility on Reddit or Hacker News is very similar to estimating position bias in search results and search ad rankings. We exploit this similarity in our techniques. 

Reddit and Hacker News also admit a clean definition of ``quality''.  Each site orders articles by a function of upvotes (or upvotes minus downvotes in Reddit's case) and age. For convenience, we'll use the term score to mean number of upvotes in the case of Hacker News, or the difference between upvotes and downvotes for Reddit.  Holding all else equal, Reddit and Hacker News implicitly consider articles with a greater score to be better than articles with a lower score. We will then define quality as the score an article would receive if all articles were shown to the same number of users,  and in the absence of confounds like social signals. The MusicLab experiment accomplished this by randomizing the ordering of songs and not displaying the number of downloads that each song has. The reality of voting is nowhere close to this process, so we must instead estimate this hypothetical quality measure from data. 

Lastly, recent empirical work shows that popularity on Reddit exhibits signs of a distorted relationship between quality and popularity. \citeauthor{gilbert2013widespread} \shortcite{gilbert2013widespread} studies popularity of images on Reddit and finds that over half of popular image submissions are actually reposts of previous submissions. The same picture may receive no upvotes on it's first submission but its second or third submission may gain thousand of upvotes.

\subsection{Our Contributions}	

The main contribution of this paper is a method for estimating article quality on Reddit and Hacker News. The key to our analysis is the use of time-series observations of voting behavior for each article. Observing the same article at different points in it's life allows us to examine the influence of different factors on voting data.

We begin by showing a small example of a factor that causes a distortion between quality and popular. We identify a set of articles on Hacker News that had identical voting patterns in the first 30 minutes of their lifetimes, and find that articles in this set that (randomly) began on the front page of Hacker News eventually became more than twice as popular than articles that began on lower pages.  

We then develop a simple poisson regression model for learning parameters from observed data, with the goal of using those parameters to estimate the score an article would have received if the voting were conducted in a bias-free manner. Our base model only includes factors for article and position effects but we expand this model to also include factors for the time decay of article quality and for potential social influence effects. Since we lack the ability to evaluate against ground truth data from Reddit or Hacker News, we evaluate this model on data from the MusicLab experiment. We find this method is effective at recovering ground truth quality parameters, and further show that it has good explanatory power for Reddit and Hacker News data. 

Using these estimates, we examine the relationship between article quality and popularity. We first verify the existence of a large position bias on Reddit and Hacker News. We then find a surprisingly strong relationship between quality and popularity, indicating that the most popular articles on Reddit and Hacker News are likely the best articles in the system.  

Lastly we expand Gilbert's study of reposting behavior on Reddit and show that reposters actually helps Reddit aggregate content that is popular on the rest of the web. Specifically, we show that the number of submissions of an article is positively correlated with its external popularity, and these increased number of postings raise the probability that at least one becomes popular.

\section{Related Work}\label{sec:related_work}

This work is mostly closely related to the literature on predicting popularity, although we emphasize the goal is different. The goal of the prediction literature is to predict popularity within real systems, which are subject to the biases that we listed in the introduction, whereas our goal is to ``predict'' popularity in a hypothetical bias-free world. One implication of \cite{salganik2006experimental} is that predicting popularity is inherently difficult because cultural markets tend to construct preferences rather than reveal them. However prediction literature shows  popularity can be predicted accurately \emph{if} we observe a small amount of early popularity. \cite{szabo2010predicting} show that Youtube views can be predicted accurately using a log-linear extrapolation of early popularity. Recent work has improved the accuracy of this method \cite{pinto2013using} but still only uses features related to early popularity. Both \cite{cheng2014can} and \cite{bakshy2011everyone} find that content features are weak features in comparison to structural features, such as the networks of original posters and the timing of early shares and retweets. Some scholars \cite{bandari2012pulse} have proposed and tested prediction methods that only use content features, but \cite{arapakis2014feasibility} finds that predicting news popularity at cold-start with only content-features provides little accuracy over a trivial baseline. This is not to say that content-only prediction is impossible but the MusicLab experiment and related experiment provide a solid theoretical grounding for why content-only prediction should be difficult.

Several scholars have used data from the MusicLab experiment to study the roles of article qualities and social influence in popularity. \citeauthor{krumme2012quantifying} \shortcite{krumme2012quantifying} uses the MusicLab data to show that social influence affects a user's choice of what songs to listen to but not their probability of downloading a song after listening to it. \citeauthor{abeliuk2014measuring} \shortcite{abeliuk2014measuring} use the model of \citeauthor{krumme2012quantifying} to estimate article quality from the MusicLab data and design a ranking algorithm to maximize the number of downloads. Similarly \cite{lerman2014leveraging} \shortcite{lerman2014leveraging} design a news aggregator experiment on Mechanical Turk to test the efficacy of different ranking algorithms, and find that ordering by popularity was best at focusing user attention on high quality articles. In follow-up work, \citeauthor{hogg2014effects} \shortcite{hogg2014effects} use the same experiment to test the effect of social influence. In doing so, they use a model to estimate article quality which is similar to the one we employ (see section \ref{sec:model}). 

Although there aren't many studies on Reddit (see below), social news aggregators have been studied in recent years. \citeauthor{lampe2004slash} \shortcite{lampe2004slash} analyze the comment moderation system on Slashdot and show that many comments never receive a fair judgment because of ``rich-get-richer'' style effects.  Lerman et al \shortcite{hogg2009stochastic,lerman2010using} studied popularity on Digg, and demonstrated that popularity prediction accuracy could be improved by tailoring a model to the algorithm and interface that Digg used. \citeauthor{hodas2014simple} \shortcite{hodas2014simple} use an explicit model of user attention and visibility to explain differences in information diffusion between Digg and Twitter. 

Finally there's a nascent literature that uses Reddit as a data source. While our approach is agnostic to article features, other scholars have used qualitative and quantitative to study popularity and preferences on Reddit. \citeauthor{lakkaraju2013s} \shortcite{lakkaraju2013s} decompose article popularity into separate effects: the quality of the content itself and the appeal of the title of the content. They show that textual features of the title play a large role in determining popularity. \citeauthor{leavitt2014upvoting} \shortcite{leavitt2014upvoting} study the behavior of a subreddit dedicated to Hurricane Sandy news and qualitatively analyze preferences of that community, and the resulting popularity of different types of stories within that subreddit.

\section{Data}\label{sec:data}

\textbf{Reddit} 
\\ Reddit is composed of many different sub-communities called ``subreddits''. For example ``r/news''\footnote{by convention, ``r/'' is prefixed to the name of a subreddit when referring to it} is the subreddit for discussing news and current events. Links must be submitted to a subreddit. Any logged-in user may either upvote or downvote any link on Reddit. When a user visits the front page of Reddit (reddit.com), they are shown articles from a combination of different subreddits that they are subscribed to. New users and users without accounts are shown articles from a combination of the ``default'' subreddits. Although the main ranking is personalized for each user,  articles are ranked the same for all users when they visit a particular subreddit (i.e. reddit.com/r/news). Within a subreddit, articles are ranked in decreasing order of their ``hot score'', which is defined by: 
\[ \log(u_i - d_i) - \frac{1}{750} \text{age}_i \]
Where $u_i, d_i$ is the number of upvotes and downvotes received by article $i$ and $\text{age}_i$ is the number of minutes between the current time and the time the article was submitted\footnote{750 is the number of minutes in 12 1/2 hours.}. There's additional logic to handle the case where $d_i \geq u_i$ but most of our observations have $u_i > d_i$. The algorithm for ranking articles on the front page is more complex and we omit these details because this work only studies dynamics within particular subreddits. 

\textbf{Hacker News}
The design of Hacker News is a bit simpler than Reddit's in a few ways. First, Hacker News allows people to upvote stories but not to downvote them. Second, there are only two different article rankings: the ``new'' ranking which is a chronological list of articles, and the ``top ranking''. In the ``top ranking'', articles are ranked according to the following ``top score'' (a term we use for the sake of convenience):
\[\frac{(u_i - 1)^{.8}}{ (age_i + 2)^{1.8}} \cdot penalties_i \]
Where $u_i$ is the number of upvotes for article $i$ and $age_i$ is the time (in hours) elapsed between submission and the current time. $penalties_i$ is a factor related to certain features of the article, such as whether it is ``light weight'' or ``controversial''. The details of penalties are not made public but in most cases the penalty factors have little impact. To appear in the top-ranking of Hacker News, an article's score must exceed some minimum threshold. The number of articles in the top ranking can vary because of the threshold. 

\textbf{Data Collection} 
\\ 
We collect data at 10 minute intervals over a two week period from 5/26/14 to 6/6/14 for each site. For Hacker News, we collect all articles in the top ranking and new ranking and record the number of upvotes, comments, and position of each article (as well as static metadata like title, author, etc). We can compute the number of votes an article received by comparing the number of votes in subsequent time periods. For our purposes, each observation is a tuple $(t, i, j, v^t_i)$, meaning that article $i$ at time $t$ was observed in position $j$, and received $v^t_i$ upvotes in the time period $t$ to $t + 1$. 

For Reddit, we collect the top 500 articles of the hot rankings for a number of different default subreddits, and record the number of upvotes, downvotes, position, and score (difference of upvotes and downvotes) for each article. Each observation is a tuple $(t, i, j, u^t_i, d^t_i)$, meaning that article $i$ at time $t$ was observed in position $j$ and received $u^t_i$ upvotes and $d^t_i$ downvotes in the time period $t$ to $t+1$. The values we use for $u^t_i$ and $d^t_i$ are actually not the raw values collected from Reddit; due to an anti-spam practice on Reddit called ``vote fuzzing'', we had to transform the data extensively. We describe this transformation in the appendix.

\section{A Simple Example}\label{sec:luck}

In this section we demonstrate a concrete example of why observed popularity may only be a distorted measure of article quality. We show that the initial position of an article within Hacker News' top-ranking has a large influence on its eventual popularity. Furthermore we show initial positioning is partially determined by random factors.


Articles only appear in the top-ranking of Hacker News when their score exceeds some minimum threshold. We define initial position as the position of an article when it first enters the top-ranking. We certainly have some measurement errors for initial positions because we scrape at 10-minute intervals, so it's possible that an article enters at some position $j$ but we first see it at position $j'$. However the median movement in position between time intervals in only 2 ranks, so we feel this measurement error has an insignificant effect. The initial position of an article is partially a function of the quality of an article. To mitigate endogeneity issues, we first select a subset of articles that are essentially the same, according to their top-score, when they enter the top-ranking of Hacker News. Specifically, we only consider articles that enter the top-ranking with exactly 3 upvotes and within 30 minutes of submission to Hacker News. Since these articles have roughly the same top-score, the variation in initial position is determined by the scores and ages of the other articles in the ranking.

\begin{figure}
\centering
\begin{subfigure}{0.4\textwidth}
\includegraphics[width=\textwidth]{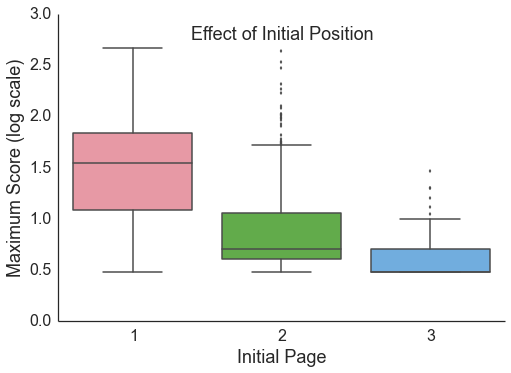}
\end{subfigure}
\caption{The effect of initial placement on the final score of articles on Hacker News.}
\label{fig:hn_initial}
\end{figure}

Hacker News breaks its ranking into different pages by every 30 positions. Figure \ref{fig:hn_initial} shows a box plot of the final score of articles, where articles are grouped by the initial page that they appear on. The median score of articles that initially appear on the first page is 35 versus 5 and 4 for the second and third page. The mean scores are 57, 20, and 4. After adding controls for time and day of submission and running a linear regression, we find that articles that initially appear on the front page receive 57 more upvotes on average. 

The reason for this difference is fairly simple. A small fraction of Hacker News users look at the second page, so stories that start there never have a chance to gain many votes. This is just one example of a dynamic that could potentially distort the relationship between quality and popularity.

\section{Model}\label{sec:model}

In this section we present and evaluate our method for estimating article quality. As discussed in the introduction, our measure of intrinsic article quality is the score an article would have received if voting were conducted in a bias-free way. Specifically, we want to estimate the score an article would have received if it were exposed to same number of randomly-chosen users, and in the absence of any social signals like the current score of the article or the number of comments. While this definition of quality is clearly not the only definition of quality, nor is it suitable for every type of article on Reddit or Hacker News (see section \ref{sec:limitations} in the appendix for a discussion of this issue), we feel its a reasonable definition.  

\subsection{Estimation}

The primary issue for estimating these parameters we don't observe the number of users who viewed an article but decided not to vote on it. If we observe that an article only received 5 votes, we don't know if 5 or 500 users viewed that article. Even if we knew the number of users who visited a page (presumably these sites do), we still couldn't be sure because articles at the top of the page may receive many more views than articles at the bottom. 

Fortunately this is a common problem encountered in estimating the click-through-rates of search results and search ads \cite{dupret2008user},\cite{chen2012position}, \cite{craswell2008experimental}. One model used in this literature is the \emph{examination hypothesis}, proposed by \cite{richardson2007predicting}, which models the probability of a user clicking on article $i$ in slot $j$ as a two-step process. With probability $\pview{j}$ a user examines the article in position $j$, independent of the article in position $j$. If the users examines position $j$, they click on that article with probability $q_i$. Then the $p$ and $p$ parameters are estimate from search log data, typically via maximum likelihood estimation. 

Direct application of this model isn't possible because the granularity of our data is votes cast over a 10 minute interval, rather than voting data at the individual level. Instead of estimating the probability that a single user votes on an article, we estimate the rate at which an articles receives votes. Recent work \cite{chen2012position} shows that the binomial model in the examination hypothesis can be replaced with a poisson model. As it turns out, their formulation is exactly equivalent to the following standard poisson regression with fixed effects for article quality and position bias:
\begin{equation*}
v^t_{i} \sim \text{Poisson}(\exp(p^t_i + q_i))
\end{equation*}

Where $v^t_i$ is the number of votes received by article $i$ at time $t$ and $p^t_i$ is the position it appeared in. The fitted $q_i$ parameters are then used to estimate the quality of each article (described in more detail in section \ref{sec:popularity} ). While the approach in \cite{chen2012position} yields an efficient closed-form solution for the MLE, the poisson regression gives us the flexibility to add other factors. We choose the poisson regression approach for this reason. 

\subsection{Time Decay and Score Effects}

The above model accounts for the primary effects of position bias and article quality but there are other factors that affect voting. The first is that article quality may decay over time, perhaps because the news is less relevant or a large fraction of users may have seen the article on earlier visits. We add an age term to account for such effects. $age_i^t$ is the time difference (in hours) between the time of observation and the submission time of article $i$.    

Both Reddit and Hacker News display the current score of each article, which gives a signal about the article. Prior work \cite{hogg2014effects}, \cite{muchnik2013social},\cite{krumme2012quantifying}, \cite{salganik2006experimental} shows these signals can have significant effects on user viewing and voting behavior. We add a term for score effects but first apply a log transformation to scores to account for the large disparities in scores on Reddit and Hacker News. Let $S^t_i$ be the score of article $i$ at time $t$. The full model is:  
\begin{equation*}
v^t_{i} \sim \text{Poisson}(\exp\{p^t_i + q_i + \beta_{age} \cdot age^t_i + \beta_{score} \cdot \log(S^t_i) \})
\end{equation*}

In summary, the full model estimates an article quality effect $q_i$ for each article, a position bias effect $p_j$ for each position, a time decay effect $\beta_{age}$, and a score effect $\beta_{score}$. We use the StatsModels python module\footnote{\url{http://statsmodels.sourceforge.net}} to implement the poisson regression, with the L-BFGS method to optimize the likelihood function \cite{nocedal1980updating}.

\section{Evaluation}

Unfortunately we lack the ability to evaluate our estimates against ground truth from Reddit or Hacker News. Instead we validate the model in two different ways. First we apply this model to data from the MusicLab experiments \cite{salganik2006experimental} and compare against the ground truth estimates from that experiment. We find that the poisson regression recovers accurate estimates of the ground truth quality data. Second we show the poisson model has good explanatory power for the Reddit and Hacker News data, even when evaluated on out-of-sample data during cross-validation. 

\subsection{MusicLab}

Participants in the MusicLab experiment \cite{salganik2006experimental} were shown a list of unknown songs that they could listen to and download. When participants entered the website, they were assigned to 1 of 9 different worlds. In the first 8 worlds, songs were ordered by the number of downloads the song received within that world (the download count was also displayed to users). In the 9th world, songs were shown in a random order to each user and the current download count was not displayed.  

We use this data to assess the accuracy of the poisson regression method. We use data from the first 8 worlds, the ones which were ranked by popularity and subject to social influence, to estimate the number of downloads of each song in the random world. We fit the following model: 
\[d^{t,w}_{i} \sim \text{Poisson}(\exp\{q_i + p^{t,w}_i + \beta_{social} \cdot D^{t,w}_i\}) \]

\begin{figure*}
\begin{subfigure}[t]{0.3\textwidth}
\includegraphics[width=\textwidth]{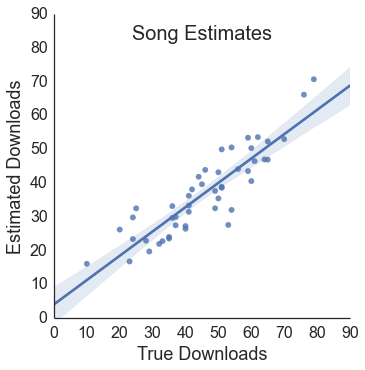}
\subcaption{Estimated versus ground truth parameters from the MusicLab experiment. Each (x,y) data point represents the actual number of downloads of a song in the random world versus the predicted number of downloads.}\label{fig:music_lab_downloads}
\end{subfigure}
\begin{subfigure}[t]{0.3\textwidth}
\includegraphics[width=\textwidth]{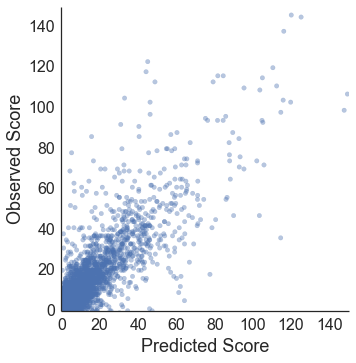}
\subcaption{Observed versus predicted score values \quad for r/pics}\label{fig:pics_predicted_observed_scatter}
\end{subfigure}
\begin{subfigure}[t]{0.37\textwidth}
\includegraphics[width=\textwidth]{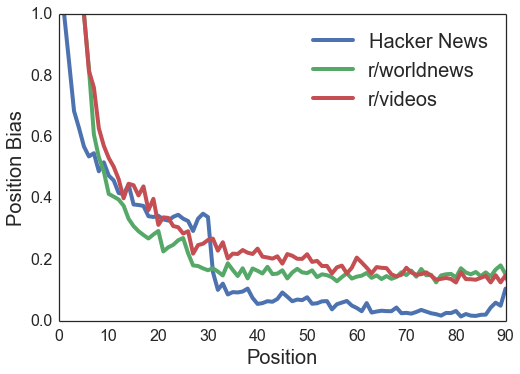}
\subcaption{Estimated position bias of top 90 positions for Hacker News and select subreddits. Position biases have been normalized such that $\pview{1} = 1$.  }\label{fig:position_bias}
\end{subfigure}
\caption{}
\end{figure*}

Where $d^{t,w}_{i}$ is an indicator variable for whether the $t^{th}$ user in world $w$ downloaded song $i$, $p^{t,w}_i$ was the position that song $i$ appeared in for that user, and $D^{t,w}_{i}$ is the number of downloads of song $i$ in world $w$ when user $t$ visited. We then use the fitted parameters to predict the expected downloads of each song in the randomly-ordered world:
\[ \hat{D}^9_i = \sum_{t=1}^T \exp \{q_i + p^{t,9}_i\} \]
The score term is removed because current downloads were not displayed in the random world. These predicted values are compared against the actual number of downloads, $D^9_i$, in figure \ref{fig:music_lab_downloads}. Although the predicted number of downloads underestimates the true downloads, it predicts the data very well up to that scalar factor. If we were to scale the download counts, the data would tightly fit the $y=x$ line. 




\subsection{Reddit and Hacker News}

We now test how well this model predicts the time-series data from Reddit and Hacker News. For each site, the predicted number of votes for each observation is the conditional mean of the poisson distribution, i.e.,  
\[ \hat{v}^t_i = \exp\{ q_i + p^t_i + \beta_{age} \cdot age^t_i + \beta_{score} \cdot \log(S^t_i) \}\]

For Reddit this only predicts the number of votes on an article, not the increase in score. We can directly estimate an article's probability of getting an upvote conditional on the article receiving any vote by the fraction of upvotes to total votes. The unconditional rate of upvoting is the rate of voting times the conditional upvote probability, and the predicted growth in score is just the upvote rate minus the downvote rate. Let $r^{up}_i$ be the observed ratio of upvotes to total vote for article $i$ and $r^{down}_i$ be the ratio of downvotes. The predicted growth in score for article $i$ at time $t$ is: 
\[ \hat{s}^t_i = \hat{v}^t_i \cdot ( r^{up}_i - r^{down}_i )  \]

We evaluate the accuracy of predictions using the coefficient of determination ($R^2$ value), mean absolute error, and mean squared error. In addition to reporting the accuracy on the in-sample data, we run a 5-fold cross-validation and report prediction accuracy on the out-of-sample data points.

\subsection{Results}
The results are shown in table \ref{table:accuracy_metrics}. The model performs well for both in-sample and out-of-sample prediction, capturing between 50\% and 80\% of the variance in the voting data. To visualize the prediction, we plot observed scores versus predicted scores for out-of-sample data points from r/pics in figure \ref{fig:pics_predicted_observed_scatter}. Each data point represents the estimated and observed value for the difference in upvotes and downvotes that article received in a 10 minute interval. While the fit is reasonably good, the data is over-dispersed. The poisson model assumes that conditional variance is equal to the conditional but this doesn't hold in our data (technically, this is not readable from the plot because we show the score, which is the difference of two poisson variables, but we verified this using the prediction of the pure number of votes as well). While this assumption on the variance isn't necessary for valid estimate of the maximum likelihood parameters, it suggests that the poisson model can be improved upon.

The predictions in table \ref{table:accuracy_metrics} were made using the full poisson model, i.e. the one that includes terms for time-decay and score effects. Table \ref{table:model_comparison} shows the average cross-validated $R^2$ values for the base poisson model, the model with just a time-decay factor, and the full model. In most cases, gains in accuracy are driven primarily by the addition of the time-decay factor but the score effects do help. We use the full model to estimate article quality except for r/news and r/pics. The values for position bias in the full poisson model behaved oddly in those two cases; the resulting estimates implied that positions 200 to 300 received more views than the top 50 positions. It seems that the position bias for those top 50 positions were ``pushed'' into the score parameter. Although the full model was marginally more accurate, we chose to drop the score term for those two datasets because of this unintuitive behavior.


\begin{table}[t]
\scalebox{.7}{
\begin{tabular}{|l|l|l|l|l|l|l|}
\hline
 & \multicolumn{3}{l|}{In Sample Predictions} & \multicolumn{3}{l|}{Out of Sample Predictions} \\ \hline
 & $R^2$ & MAE & MSE & $R^2$ & MAE & MSE \\ \hline
r/pics* & 0.76 & 1.09 & 7.30 & 0.62 (0.01) & 1.14 (0.01) & 8.51 (0.40) \\ \hline
r/videos & 0.79 & 1.15 & 9.62 & 0.65 (0.03) & 1.22 (0.01) & 13.64 (2.59) \\ \hline
r/todayilearned & 0.71 & 1.75 & 22.66 & 0.61 (0.03) & 1.85 (0.02) & 32.24 (3.74) \\ \hline
r/news* & 0.56 & 1.11 & 3.63 & 0.57 (0.01) & 1.14 (0.01) & 3.87 (0.18) \\ \hline
r/worldnews & 0.57 & 1.27 & 9.10 & 0.52 (0.01) & 1.32 (0.01) & 10.65 (1.17) \\ \hline
Hacker News & 0.69 & 0.70 & 1.82 & 0.65 (0.01) & 0.74 (0.01) & 2.08 (0.11) \\ \hline
\end{tabular}}

\caption{Accuracy metrics for the full Poisson model. In sample value are trained on and predicted for the same dataset. Out-of-sample are trained on a train set and predicted for a test over 5 fold cross-validation. }
\label{table:accuracy_metrics}
\end{table}

\begin{table}[h]
\centering
\begin{tabular}{|l|l|l|l|}
\hline
                & Base & Base + Time & Full \\ \hline
r/pics$^*$          & 0.56 & 0.58        & 0.62 \\ \hline
r/news$^*$          & 0.53 & 0.55        & 0.59 \\ \hline
r/worldnews     & 0.51 & 0.51        & 0.52 \\ \hline
r/todayilearned & 0.61 & 0.59        & 0.61 \\ \hline
r/videos        & 0.63 & 0.58        & 0.65 \\ \hline
Hacker News     & 0.51 & 0.63        & 0.65 \\ \hline
\end{tabular}
\caption{Average $R^2$ values over cross-fold validation for the three models. The starred subreddits indicate that the full model is not used for quality estimation because of bad fitted values for position effects.}
\label{table:model_comparison}
\end{table}

\section{Analysis}\label{sec:popularity}

We first use these estimates to quantify position bias on Reddit and Hacker News. Figure \ref{fig:position_bias} shows the relative view rates for the top 90 positions. We only show data for two subreddits but the trends hold on each subreddit studied.  The relative view rate for position $j$ by is computed by exponentiating the fitted $p_j$ parameter from the poisson regression and scaling so the maximum view rate in a subreddit is equal to 1. The curve for each subreddit begins at position 5 because we discard observations from the top 5 positions of each subreddit (see the appendix for the reasoning behind this). Each dataset shows an exponential decline in view rate but Hacker News has a particularly sharp drop at its page break (position 30 to 31), whereas the subreddits display a smoother decline. The general shape of position bias is consistent with estimates from other platforms \cite{krumme2012quantifying},\cite{lerman2014leveraging}. 

\subsection{Quality and Popularity}

We now measure the relationship between estimated quality and popularity on Reddit and Hacker News. An article's quality is defined as the the expected score article $i$ would receive if all articles in that subreddit were shown to the same number of users. For convenience, we also scale qualities such that the maximum quality article in a given subreddit is equal to 1. Given the fitted $q_i$ parameters from the poisson model, the estimated quality for a Hacker News article is:
\[Q_i = \frac{e^{q_i}}{\max_j e^{q_j}} \]
For Reddit, estimated quality is: 
\[ Q_i = \frac{e^{q_i} \cdot ( r^{up}_i - r^{down}_i)  }{\max_j e^{q_j} \cdot (r^{up}_j - r^{down}_j )}\]
where $r^{up}_i$ is the observed fraction of upvotes to total votes by article $i$ ($r^{down}_i$ defined similarly). 

\begin{figure*}
\begin{subfigure}[t]{0.33\textwidth}
\includegraphics[width=\textwidth]{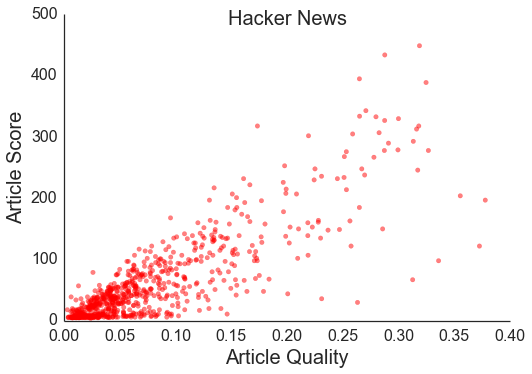}
\subcaption{Observed popularity versus estimated quality for Hacker News. X-axis is truncated for visualization purposes but only a few data points were omitted. }\label{fig:hn_score_quality}
\end{subfigure}
\begin{subfigure}[t]{0.33\textwidth}
\includegraphics[width=\textwidth]{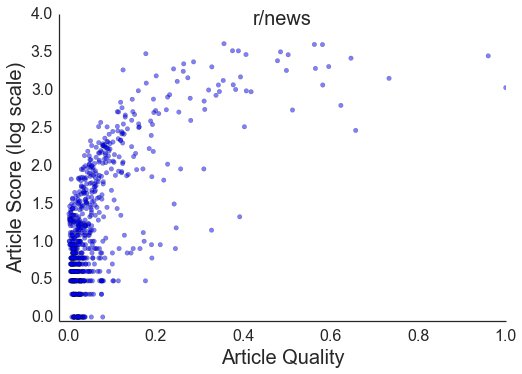}
\subcaption{Observed popularity versus estimated quality for r/news}\label{fig:news_quality_score}
\end{subfigure}
\begin{subfigure}[t]{0.33\textwidth}
\includegraphics[width=\textwidth]{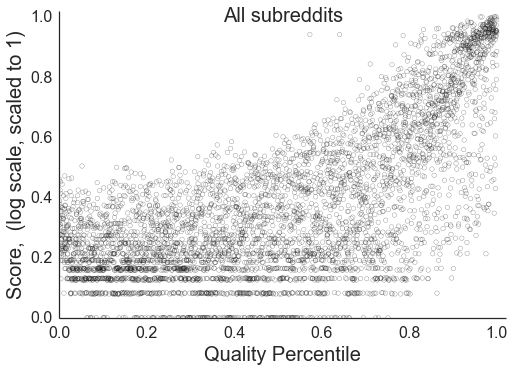}
\subcaption{Popularity versus quality for all subreddits. Scores are first log-transformed and then scaled by the maximum score in the subreddit. Qualities are measured by quantile.}\label{fig:reddit_quality_score_rank}
\end{subfigure}
\label{fig:quality_score}
\caption{A sample of popularity versus estimated quality plots for Hacker News and Reddit.}
\end{figure*}

Figures \ref{fig:hn_score_quality},\ref{fig:news_quality_score}, and \ref{fig:reddit_quality_score_rank} show a few examples of the relationship between quality and score. Hacker News, shown in figure \ref{fig:hn_score_quality} has the highest correlation between score and quality, while r/news, shown in figure \ref{fig:news_quality_score}, has one of the weakest relationships. Figure \ref{fig:reddit_quality_score_rank} shows the relationship for all subreddits. In order to compress everything into one plot, we use the quantile of article quality (an article with a quantile of .75 has a higher quality than 75\% of other articles in the same subreddit). Second, we log-transform the scores of each article, and then scale by the maximum log-transformed score within the subreddit. 


The relationship between quality and popularity is consistent with expectations from the MusicLab experiments. Popularity is generally increasing with quality but articles of similar quality can have large differences in popularity. However we find that there are few instances of a mediocre quality article becoming one of the most popular articles in a subreddit, and few instances of high quality articles ending up with low scores. In general, the relationship between popularity and quality is stronger on Reddit and Hacker News than the MusicLab experiment. The first column of table \ref{table:spearman} lists the spearman correlation coefficients between quality and popularity. Hacker News has the strongest relationship with a correlation of .8 and r/worldnews has the weakest with a correlation of .54.

We had initially expected the quality-popularity relationship to be weaker on Hacker News than Reddit because of the lack of the downvote. Our theory was that a low quality article that made it to the front page of Hacker News would remain for a long time and get popular because there was no ability to downvote it off. This theory is partially true; the second column in table \ref{table:spearman} shows the relationship between quality and total views. We estimate total views by $\sum_t \exp\{p^t_i\}$, i.e. the sum of position biases for the positions that article $i$ appeared in during its lifetime. The relationship between total views on Hacker News is much weaker than on Reddit, indicating that lower quality articles are being seen comparatively more often on Hacker News. However this did not translate to a weakened quality-popularity relationship as we had expected.

\subsection{Discussion}

There is one important caveat to these results. Many articles submitted to Reddit and Hacker News fail to gain any votes and quickly disappear. For example, there were 5000 articles submitted to Hacker News over the period of observation but only 1500 of them even made it to the top ranking. These ignored articles did not generate enough observations to be included in our dataset. So when we state that the relationship between quality and popularity is fairly strong, we must interpret that as only being among a set of articles that received at least a reasonable amount of attention. In the Reddit dataset, the median article received 38 votes (upvotes plus downvotes), while the median Hacker News article received 21 votes, with a minimum of 3 votes in each case. Its likely there are a number of high quality articles that were discarded from this study because they didn't generate enough observations. Developing methods to handle these cases is an interesting direction for future work. 

With this caveat in mind, these results imply that relative popularity is a good indicator of relative quality among articles that received a reasonable amount of attention. We now return to the thought experiment proposed in the introduction: If we asked a large number of people to vote on every music video on Youtube, how would ``Gangnam Style'' fare in that election? Based off of this work, we expect that it would rank higher than most other highly-watched videos. However its completely possible, and perhaps even likely, that a relatively unknown video would claim the top spot.

\begin{table}[t]
\centering
\begin{tabular}{|l|l|l|}
\hline
 & Score & Views \\ \hline
Hacker News & .80 & 0.49 \\ \hline
r/todayilearned & .75 & 0.81 \\ \hline
r/videos & .63 & 0.70 \\ \hline
r/worldnews & .54 & 0.70 \\ \hline
r/news & .59 & 0.75 \\ \hline
r/pics & .63 & 0.77 \\ \hline
MusicLab & .57 & 0.35 \\ \hline
\end{tabular}
\caption{Spearman correlation between estimated quality and observed score, and quality and estimated views.}
\label{table:spearman}
\end{table}

\section{Reposts}

\newcommand \numVideos {61,110 }
\newcommand \numSubmissions{91,841 }
\newcommand \numRepostedVids{11,297 }
\newcommand \numReposts{42,028 }

As discussed in the last section, many articles on Reddit or Hacker News go almost completely ignored. A recent estimate shows that over half of links on Reddit receive at most 1 upvote \cite{olsonreddit2014}. The work of \citeauthor{gilbert2013widespread} \shortcite{gilbert2013widespread} shows that it isn't because this content is necessarily bad; Gilbert finds that over half of popular images on Reddit were submitted and ignored a few times before they became popular. It seems problematic for Reddit's role as an aggregator of the most interesting content on the web.

One subtle point of \cite{gilbert2013widespread} is that those images eventually became popular, even if it took a few reposts. Although Reddit's voting mechanism failed to popularize some good content, the reposting behavior of Redditors corrected this failure. In this section we briefly explore the role of reposts in popularizing good content on Reddit. We find evidence that the number of reposts of an article is positively correlated with it's quality. Unfortunately we cannot use the methods from previous sections to estimate quality because the scope of our time-series data is too limited to capture much reposting behavior. Instead we limit ourselves to Youtube videos submitted to Reddit and use Youtube views as a proxy for quality.

\begin{figure*}
\centering
\begin{subfigure}{0.4\textwidth}
\centering
\includegraphics[width=\textwidth]{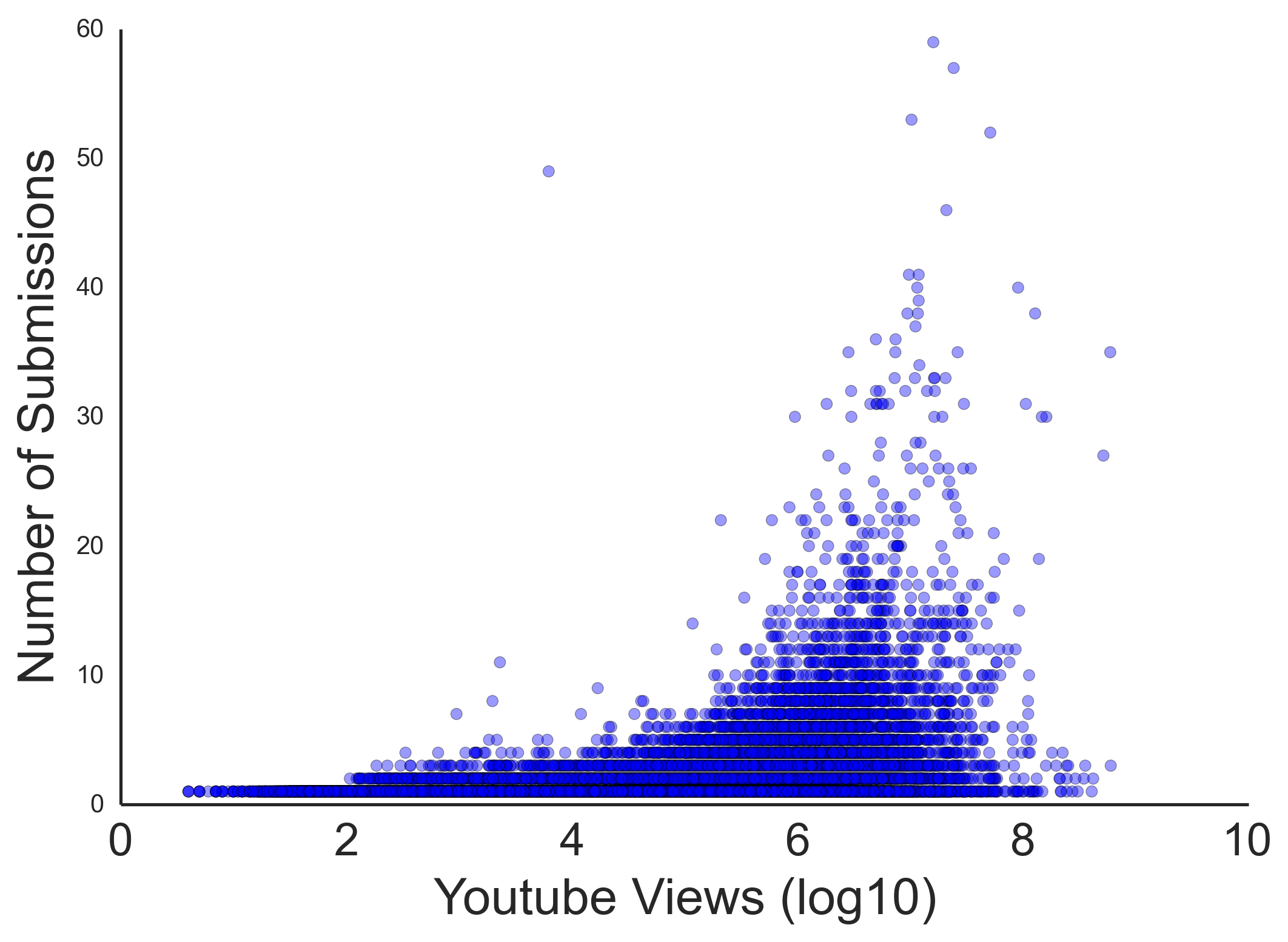}
\subcaption{Relationship between Youtube views and submissions to r/videos in 2012}
\label{fig:youtubeViews_Submissions}
\end{subfigure}
\begin{subfigure}{0.4\textwidth}
\centering
\includegraphics[width=\textwidth]{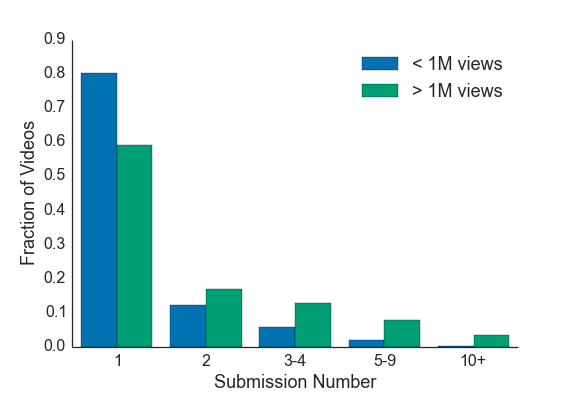}
\subcaption{Videos are grouped by the submission number of their highest scoring submission}
\label{fig:reposts_barchart}
\end{subfigure}
\caption{}
\end{figure*}

We study all videos that were uploaded to Youtube and submitted to r/videos in 2012. We're left with a set of \numVideos unique videos after removing videos we were unable to retrieve metadata for. These videos were submitted a total of \numSubmissions times to Reddit; \numRepostedVids of these videos were submitted multiple times, generating a total of \numReposts reposts to Reddit. Figure \ref{fig:youtubeViews_Submissions} shows a scatter plot of number of posts to Reddit versus Youtube views for each video. There's a strong positive relationship between views and submissions, suggesting that users submit popular Youtube videos more frequently. Videos with more than 1 million views were submitted twice as often to Reddit; the mean and median number of submissions for all videos are 1.5 and 1 while the mean and median for videos with more than one million views are 3.6 and 2.

These reposts are actually responsible for surfacing many Youtube videos that would have gone unnoticed on Reddit otherwise. Figure \ref{fig:reposts_barchart} shows a bar chart of popular posts on r/videos, where popular is defined as being in the top 10\% of posts in 2012 as measured by score \footnote{This equates to having a score of 23 or greater.}. We further divide videos by whether they have more or less than 1 million Youtube views. Videos are grouped by the number of posts until it became popular on Reddit. Videos in the first bucket were popular on their first submission, videos in the second bucket were popular on their second, etc. This plot demonstrates that only 59\% of videos with more 1 million views become popular on their first submission, while 76\% of videos with less than 1 million views became popular on their first submission. This difference is likely caused by the fact that more popular videos were submitted more times; we suspect that if videos with less than 1 million views were submitted as often, then these plots would be more equal. This conclusion, that reposts help popularize many videos, is similar to the conclusion of \cite{gilbert2013widespread} but our analysis further shows that reposts are particularly instrumental in popularizing videos that are externally popular.\footnote{We cannot rule out the possibility that number of submissions to Reddit is causing a rise in Youtube views but this seems unlikely given the relative size of Reddit versus Youtube in 2012.}

\section{Conclusion and Future Work}

This paper tries to understand the relationship between intrinsic article quality and popularity in social news aggregators. The heart of the problem is estimating parameters from data that allow us to reason counter-factually about the popularity of an article if the voting process were not subject to the large biases that exist in reality. We found that the most popular content on Reddit and Hacker News were, for the most part, higher quality articles than less popular content, which is surprising given the number of confounds on Reddit and Hacker News. To the best of our knowledge, this is a novel problem within the field of popularity studies. 

The poisson regression model presented in this paper is an initial approach to quality estimation, and can be improved in many ways. The most immediate is expanding the model to include a richer set of temporal features and social influence related features, such as commenting data. Although the role of social networks is relatively minimized on social news aggregators, we suspect that we could improve prediction accuracy on voting data from an article's early lifetime by incorporating such features. 

There are a number of limitations to this study, and we outline some of the more technical limitations in section \ref{sec:limitations} of the appendix. The main limitation is that our method cannot estimate the quality of a large set of articles because they do not remain in the rankings of Reddit or Hacker News long enough to generate many observations. This highlights the interesting property that early voting has a huge influence on eventual popularity. If an article receives a number of early downvotes, it effectively denies the community the chance to vote on that article. Quantifying the influence of early voters on popularity and its implications is an interesting direction for future research.

\section{Acknowledgments}

Thanks to Georgios Zervas and Brian Keegan for many helpful conversations throughout the course of this work. Thanks to participants of the Berkman Center Peer Production reading group for their insights on reposting behavior on Reddit. Thanks to Randal Olson for providing data for all Youtube videos submitted to Reddit. All error, typos, and objectionable content are our own. 

\bibliographystyle{aaai}
\bibliography{references}

\appendix

\section{Limitations}

\subsection{Model Limitations}
One assumption of the model is that the quality of an article is fixed quantity. This assumption is clearly not appropriate when the comments on an article fundamentally alter its quality. For example, users of r/AskReddit post discussion questions like ``What's the coolest thing I can buy for under \$25?'' The quality of such a thread is then highly dependent on the number of comments and quality of the comments themselves on that thread. While it may be possible to construct a structured model to capture such effects, the model we present in this paper will not do that. We purposely exclude discussion-dedicated subreddits and any post that does not redirect to an article external to Reddit or Hacker News but our data may contain some articles which are primarily just discussion threads. 

Another way in which this assumption may be problematic is that an article may be exposed to different populations of users throughout its lifetime. For example, an article in position 50 in r/news is likely only seen by users who are visiting r/news. However if that post were to reach position 1 in r/news, then it may appear on the front page for all of Reddit, resulting a different population of users voting on it. Our current model implicitly assumes that the population of voters in each case is same. We limit this issue by discarding observations of articles when they are likely to have appeared on the front page of Reddit but we do not presume to have eliminated this issue entirely.  

Finally our model assumes that the position parameters are fixed over time. Obviously there are more people viewing Reddit or Hacker News on Monday mornings than Saturday nights but our model doesn't take this into account. We attempted to add in such effects but we found that it increased over-fitting without yielding a noticeable gain in model accuracy. Instead then we limit our data to observations of Reddit and Hacker News on weekdays between 6 am and 8 pm EST. We leave it as future work to improve the model to account for such time effects. \label{sec:limitations}

\section{Data Issues}\label{sec:detailed_data}

As with any study, the raw data we collect is noisy and error prone. We address some data integrity concerns below.

\subsection{Data Granularity}
One concern is that articles may actually move positions within that time interval, so although our data says that an article appears in slot $j$, it may actually appear in slot $j'$ for a large portion of time. While this problem is unavoidable given the nature of data, we find that the change in the ranking of an article between times $t$ and $t+1$ is typically quite small. In 50\% of our Reddit observations, the position of an article was only a single position away from their previous position, and in 83\% of observations were within 5 slots of their previous positions. On Hacker News, over 90\% of observations had moved less than 3 slots. 

\subsection{Observation Inclusion Criteria}

\begin{enumerate}

\item Data must have been observed between 6am and 8pm EST on a weekday.

\item For Reddit, we limit observations to only include positions in a certain range of $[p_{min}, p_{max}]$. $p_{min}$ is defined to be 5 for all subreddits, except for r/pics where $p_{min}$ is 15. We do this to avoid observations of an article that also appeared on or near the front page of Reddit. We define $p_{max}$ to be median of the distribution of article's initial positions within a subreddit. 

\item We discard observations of articles when they are older than 12 hours. Since our model accounts for time decay, this is primarily to reduce the size of the dataset. After 12 hours, over 95\% of articles have received over 90\% of votes that they will ever receive. 

\item After removing data according to the above criteria, we finally discard any article that we don't have at least 5 observations for. 

\end{enumerate}

\subsection{Vote Fuzzing}

In the past Reddit used a practice known as ``vote fuzzing'' as a measure to combat spam and manipulation. Reddit displayed the upvotes, downvotes, and score (difference between upvotes and downvotes) but the upvotes and downvotes would each be ``fuzzed'' by the amount. This keeps the score accurate but changes the ratio of upvotes to total votes and other metrics. As of June 18, 2014 this process was stopped \footnote{\url{http://www.reddit.com/28hjga}}. Reddit no longer displays the individual number of upvotes and downvotes, and instead displays the score and the ratio of upvotes to total votes for each article. They claim the ratio and score are fairly accurate.

Our data was primarily collected in the periods before the change but we were able to use this change to retroactively ``de-fuzz'' the observed upvotes and downvotes. Since Reddit is now displaying the true score, $s^{true}$ and true ratio $\tau^{true}$, one can easily recover the true number of upvotes and downvotes. 

We cannot recompute the true values for our time-series data because we can only query retrieve the $s^{true}$ and $\tau^{true}$ for articles as they stand right now, not as they were at some time back in May. Instead, we use this information to ``reverse engineer'' the vote fuzzing method. We take advantage of the fact that articles on Reddit receive very little activity after a few days of being posted. Thus the state of an article in our collected data after 48 hours is very close to the state of the article as it would a few months later. In August 2014, we retrieved the current $s^{true}, \tau^{true}$ for these articles and used those values to calculate $u^{true}$ and $d^{true}$. 

We used this data to train a random forest regressor\footnote{We used the implementation from the scikit-learn Python module \cite{scikit-learn}.} to predict on the following to predict $u^{true}$ using $u^{obs}, s^{obs}, r^{obs}$ as features, where $(u^{obs}, s^{obs}, r^{obs})$ are the observed upvotes, score, and upvote ratio at the time we scraped the data. This method is quite accurate (average $r^2 = .96$ with 10 fold cross validation). We then use this regressor to generate the ``true'' ups and down for all data we collected. We emphasize that while this is not the ``true'' data, this method is far more accurate than using the ``fuzzed'' votes Reddit displayed prior to this change. Vote fuzzing appears to have inflated the number of votes observed at the upper tail of the distribution. This observation is consistent with anecdotal evidence from Reddit users, moderators, and administrators. 

As a final note, collecting voting data at frequent intervals is now considerably more difficult because Reddit has since changed their API. The ratio of upvotes to total votes isn't available when retrieving information in batch, only when retrieving the information for a single article. So instead of retrieving information about 1000 articles 1 API call, it now requires 1000 API calls. Collecting that information at regular intervals is impossible to do while respecting their rate limits. 




\end{document}